# APC $Nb_3Sn$ superconductors based on internal oxidation of Nb-Ta-Hf alloys

X Xu[1], X Peng[2], F Wan[1], J Rochester[3], G Bradford[4], J Jaroszynski[4] and M Sumption[3]

[1] Fermi National Accelerator Laboratory, Batavia, IL 60510, U.S.A

[2] Hyper Tech Research Incorporated, 539 Industrial Mile Road, Columbus, OH 43228, U.S.A

[3] Ohio State University, Columbus, OH 43210, U.S.A

[4] National High Magnetic Field Laboratory, Florida State University, Tallahassee, FL 32310, U.S.A

E-mail: xxu@fnal.gov

**Abstract**

In the last few years, a new type of $Nb_3Sn$ superconducting composite, containing a high density of artificial pinning centers (APC) generated via an internal oxidation approach, has demonstrated a significantly superior performance relative to present, state-of-the-art commercial $Nb_3Sn$ conductors. This was achieved via the internal oxidation of Nb-4at.%Ta-1at.%Zr alloy. On the other hand, our recent studies have shown that internal oxidation of Nb-Ta-Hf alloys can also lead to dramatic improvements in $Nb_3Sn$ performance. In this work we follow up this latter approach, fabricating a 61-stack APC wire based on the internal oxidation of Nb-4at.%Ta-1at.%Hf alloy, and compare its critical current density ($J_c$) and irreversibility field ($B_{irr}$) with APC wires made using Nb-4at.%Ta-1at.%Zr. A second goal of this work was to improve the filamentary design of APC wires in order to improve their wire quality and electromagnetic stability. Our new modifications have led to significantly improved RRR and stability in the conductors, while still keeping non-Cu $J_c$ at or above the FCC $J_c$ specification. Further improvement via optimization of the wire recipe and design is ongoing. Finally, additional work needed to make APC conductors ready for applications in magnets is discussed.

## 1. Introduction

After a steady increase for about three decades since the early 1970s (when the first $Nb_3Sn$ superconducting wires were produced), the critical current density ($J_c$) of $Nb_3Sn$ conductors stagnated [1-3]. However, efforts in the community aiming to further improve $Nb_3Sn$ $J_c$ never truly ceased because $J_c$ is such a critical parameter for magnet development, especially if the goal is not to build a small number of magnet demonstrators but to produce a large number of magnets for a collider machine, for which cost is one of the most important considerations. To build a magnet with a specific field target, using conductors with a low $J_c$ would require a larger coil size and higher conductor amount. Take, for example, the Future Circular Collider (FCC)-hh, proposed to succeed the Large Hadron Collider (LHC) [4,5]. A specification for the $Nb_3Sn$ conductor non-Cu $J_c$, which at 4.2 K and 16 T is at least 1500 A/mm$^2$, was determined based on the optimal coil current density ($J_{coil}$) for the 16 T dipoles [6,7]. The $J_c$ of the state-of-the-art $Nb_3Sn$ conductors, which are of the restacked-rod-process (RRP®) type, can meet the High-Luminosity LHC (HL-LHC) specification with a 3% margin [8]. However, the FCC $J_c$ specification is 50% above the HL-LHC specification [4-7,9]. As a result of this $J_c$ difference, it was shown by the magnet design studies for the 16 T dipoles [9] that the width of a cosine-theta coil using conductors that merely meet the HL-LHC specification is ~40% larger than that using conductors meeting the FCC specification. Thus, building the magnets using conductors only meeting the HL-LHC specification would lead to significantly higher amount of required conductors and associated magnet cost. Clearly, a significant improvement of the $J_c$ of $Nb_3Sn$ conductors is critical for building cost-effective magnets for future energy-frontier circular colliders. This will also benefit other high-field applications that use $Nb_3Sn$ conductors, such as the magnets for Nuclear Magnetic Resonance (NMR) spectroscopy.

The stagnation of $Nb_3Sn$ non-Cu $J_c$ since the early 2000s was finally lifted in 2019, achieved by a new type of $Nb_3Sn$ conductor containing a high density of artificial pinning centers (APC) formed by the internal oxidation method [10]. Although this method was used in $Nb_3Sn$ tapes in the 1960s [11], efforts to transfer it to $Nb_3Sn$ wires showed that this was a difficult task [12]. It was not until 2014 when its application in $Nb_3Sn$ wires was successfully demonstrated (first in the form of binary monofilaments) [13]. We showed in [13] that internal oxidation can be realized by making two modifications to an $Nb_3Sn$ subelement: (1) the commonly-used Nb or Nb-Ta alloy (Ta is a dopant for $Nb_3Sn$ to improve its irreversibility field, $B_{irr}$) is replaced by Nb-Zr alloy, and (2) an oxide that can be reduced by Nb and thus supply O to Nb [3] is added into the subelement, which must have a properly-designed structure such that during heat treatment the Nb alloy is able to take O from the oxide, leading to formation of fine $ZrO_2$ particles in the $Nb_3Sn$. Such fine particles enhance the flux pinning force ($F_p$) via two mechanisms. First, they refine the $Nb_3Sn$ grain size, which leads to more grain boundaries acting as flux pinning centers. Second, they directly serve as flux pinning centers – in fact, as they are point pinners, apart from enhancing the maximum pinning force ($F_{p,max}$) they also cause the $F_p$-$B$ curve peak to shift to higher fields [14]. This $F_p$-$B$ curve peak shift leads to a flatter $J_c$-$B$ curve, which enhances $J_c$ at high fields (e.g., above 10 T) but reduces $J_c$ at low fields (e.g., below 5 T) [10,13]. After demonstrating the implementation of the internal oxidation method in monofilaments [13], we then proposed to apply it to the powder-in-tube (PIT) design in order to make multi-filamentary wires [15]. The development of multifilamentary APC wires was achieved first for binary wires using Nb-1at.%Zr in 2015-2017 [16,17], then for ternary wires (with Ta doping) beginning in late 2017 [18]. Subsequently, by improving the conductor recipe and quality we were able to improve the performance of APC wires rapidly and push the non-Cu $J_c$ to the level of the FCC specification in 2019 [10]. Meanwhile, following our successful development of APC wires and demonstration of enhanced flux pinning, several other groups also began to actively study this method, each leading to some exciting discoveries [19-22]. In particular, Balachandran et al., when studying the effect of oxidation for $Nb_3Sn$ with various Nb alloys, found that the use of Nb-Ta-Hf alloys without adding O could also lead to smaller grain size than using the common Nb-Ta alloy [22]. Other effects of Ta and Hf co-doping were investigated in a subsequent study [23].

On the other hand, Zr is not the only element that can be used as solute in Nb alloys for making internal-oxidation-type $Nb_3Sn$ wires. In our previous work [24] we identified four promising elements for this purpose: Al, Ti, Zr, and Hf, as they have considerable solubility in Nb and also have a much higher affinity to O than Nb does, which is the prerequisite for internal oxidation. Among these candidates Nb-Ti, Nb-Zr, and Nb-Hf alloys had already been demonstrated to be suitable for fabricating $Nb_3Sn$ wires by the studies in the 1980s for ternary doping [25,26]. In order to see which element would lead to the strongest enhancement of $Nb_3Sn$ performance, we fabricated a mono-filamentary wire using Nb-1.5at.%Ti as well as multi-filamentary (48/61-stack) wires using Nb-4at.%Ta-1at.%Zr and Nb-4at.%Ta-1at.%Hf alloys, with all wires given sufficient O to internally oxidize the Nb alloys [24]. The results showed that internal oxidation of Nb-Ti formed large $TiO_2$ particles (from tens to hundreds of nanometers), which could not effectively serve as flux pinning centers or refine $Nb_3Sn$ grain size. Internal oxidation of Nb-4at.%Ta-1at.%Hf formed dense $HfO_2$ particles with sizes mostly below 5 nm, smaller than the $ZrO_2$ particles formed by internal oxidation of Nb-4at.%Ta-1at.%Zr (mostly 5-10 nm), both for a heat treatment temperature of 700°C. Such $HfO_2$ particles led to slightly stronger $Nb_3Sn$ grain refinement and $F_p$-$B$ curve peak shift than $ZrO_2$ particles did. In this work we present the transport properties of a recent APC wire based on internal oxidation of Nb-4at.%Ta-1at.%Hf, and compare them with those of a recent APC wire based on Nb-4at.%Ta-1at.%Zr.

It is also worth pointing out that for these two APC wires more conservative recipes were used relative to those wires made previously. In 2019 our goal was mainly to push the non-Cu $J_c$ to the level of the FCC specification, so aggressive recipes were used, which led to relatively low residual resistivity ratios (RRR) and poor electromagnetic stability [10]. One of the goals for this work is to see if more conservative recipes can lead to better wire quality and improved RRR, and to see how this affects the conductor stability and non-Cu $J_c$ (as a conservative recipe will lead to lower $Nb_3Sn$ fraction in the filaments).

## 2. Experimental

### *2.1. Samples*

Two APC wires were fabricated at Hyper Tech Research Inc. based on the internal oxidation method and the PIT filament design. Both had a 48/61 design (i.e., 48 $Nb_3Sn$ filaments and 13 Cu rods) with a Cu/non-Cu ratio of 1.15-1.25. The first wire used a tube with nominal composition of Nb-4at.%Ta-1at.%Hf (i.e., Nb-7.5wt.%Ta-2wt.%Hf) and is named "IO-Hf" here. The second wire used a tube with nominal composition of Nb-4at.%Ta-1at.%Zr (i.e., Nb-7.5wt.%Ta-1wt.%Zr) and is named “IO-Zr”. Both conductors used mixtures of Sn, Cu, and $SnO_2$ powders and had Cu/Sn and O/Nb ratios similar to a previous APC wire using Nb-4at.%Ta-1at.%Zr made in 2019 (detailed information can be found in [10], in which it was named “IO2”). Due to the more conservative recipes, the IO-Hf and IO-Zr wires have larger Nb/Sn ratios than that of the 2019 APC wire. All of these wires were drawn to 0.7 mm diameter without any breakage. Straight segments were heat treated under vacuum, all with a ramp rate of 50°C/h without any intermediate steps. The 2019 APC wire was reacted at 700°C for 70 hours, which was long enough to make sure it was fully reacted. The IO-Hf wire was reacted at 705°C for 65 hours – a few extra degrees were added to speed reaction to meet schedule, but it was not anticipated that the basic properties would depend much on a 5°C increment. More heat treatment studies were done for the more recent IO-Zr wire, and it was found that 700°C/60h was enough to fully react it. It is worth mentioning that the heat treatments used in this work may not represent the optimal schedules. For example, after the testing at the NHMFL we did further heat treatment studies on the IO-Hf wire and found that 705°C/65h led to an over-reaction and a degradation of RRR. Apart from the above APC wires, an RRP® wire for the HL-LHC project was used as a reference sample for this study. It has a wire diameter of 0.85 mm, 108 $Nb_3Sn$ subelements, and a Cu/non-Cu ratio of 1.2. It underwent a recommended heat treatment of 210°C/48h + 400°C/48h + 665°C/75h with a ramp rate of 25°C/h. More details and properties can be found in [18].  Scanning electron microscopy (SEM) images of the three APC wires after reaction are shown in Figure 1, with the (a) – (c) showing the whole filamentary regions, and (d)-(f) showing some local regions at higher magnifications so that the different components can be seen more clearly.

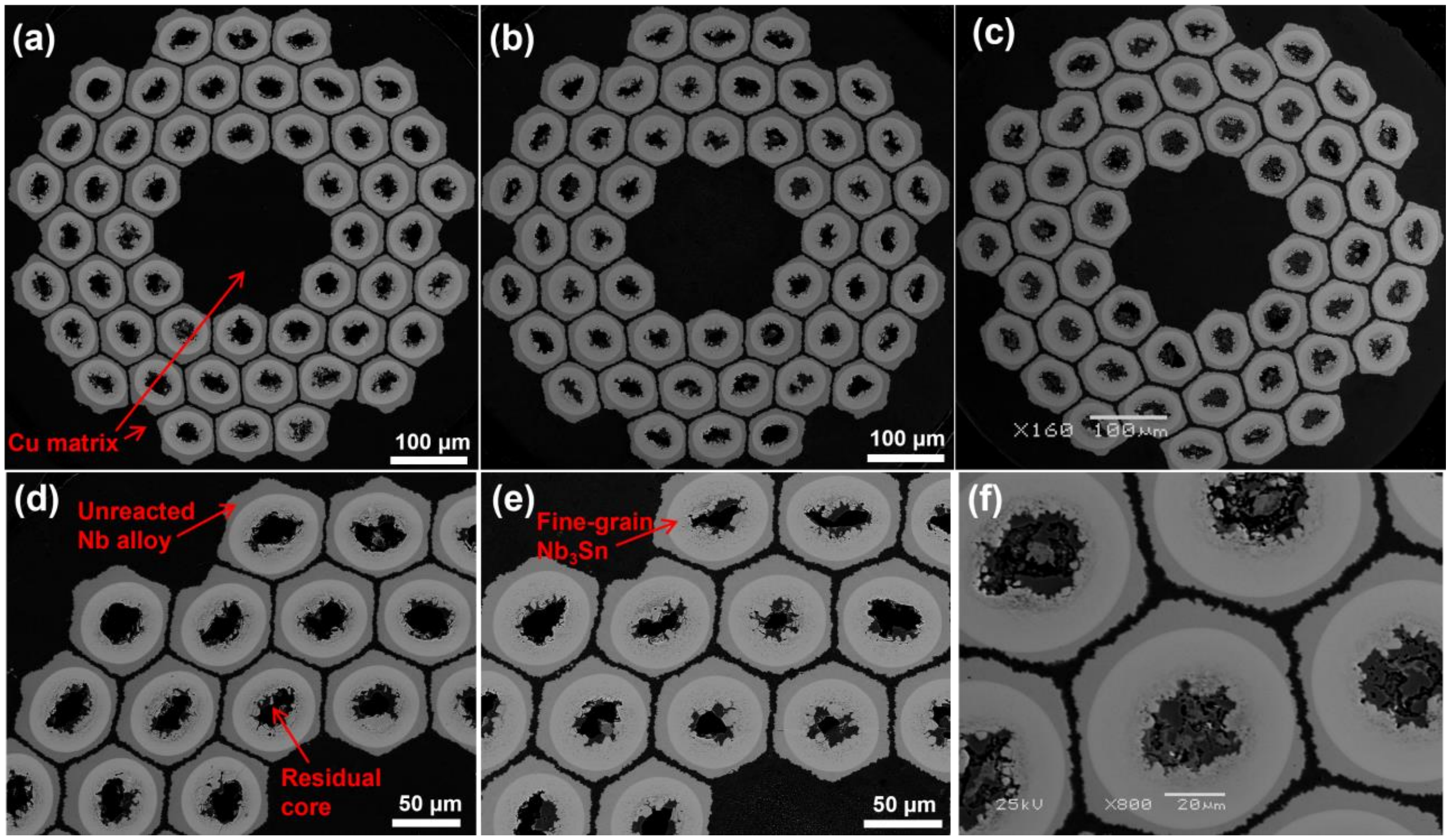

Figure 1. SEM images of (a) and (d) IO-Hf-705°C/65h, (b) and (e) IO-Zr-700°C/60h, (c) and (f) the 2019 APC wire after 700°C/70h reaction, with the (a)-(c) showing the whole filamentary regions, and the (d)-(f) showing some local regions at higher magnifications.

### *2.2. Measurements*

The voltage versus current (*V-I*) curves of the samples were measured at 4.2 K up to 26 T in a resistive DC magnet in the National High Magnetic Field Laboratory (NHMFL). Tests were performed on straight samples with the magnetic field perpendicular to the wire length; each segment was 35 mm in length with a voltage tap separation of 5 mm. A criterion of 0.1 μV/cm was used to determine the critical current ($I_c$). The segments for the *V-I* tests were later used for RRR measurements at zero field in Fermilab; for these RRR measurements a current of 1 A was used, and the voltage tap separation was 5 mm. The RRR value was taken as the resistance at room temperature (about 296 K) over that at 20 K. The irreversibility field ($B_{irr}$) and the upper critical field ($B_{c2}$) values were obtained from the resistance vs field ($R$-$B$) curves that were also measured in the NHMFL; the sample length was 15 mm and the voltage tap separation was 5 mm, with a sensing current of 0.1 A. Five samples could be measured together in each run (with the magnetic field again perpendicular to the wire length), so we always included a standard sample (e.g., RRP®) as a reference. For each measurement great care was taken to ensure that all of the samples were within the uniform-field region in the magnet.

**3. Results**

The measured non-Cu $J_c$ values of the APC wires are shown in Figure 2, along with those for the RRP® wire [18]. The engineering current density ($J_e$) values of these conductors can be calculated by multiplying their non-Cu $J_c$ values with their non-Cu fractions. Also shown in Figure 2 is the FCC $J_c$ specification. Note that here the "FCC $J_c$ specification" does not mean a single $J_c$ value at 16 T, but a $J_c$-$B$ curve generated using parameters given in [7]. The reason to use a $J_c$-$B$ curve is that for the design of a 16 T dipole, not only the $J_c$ at 16 T matters, but the $J_c$-$B$ curve in the 16-19 T range is also relevant due to the operational margin along the load line (e.g., 14% used for the FCC 16 T dipole design [4,7]) and the fact that the highest field on the conductors is typically a few percent higher than the bore field.

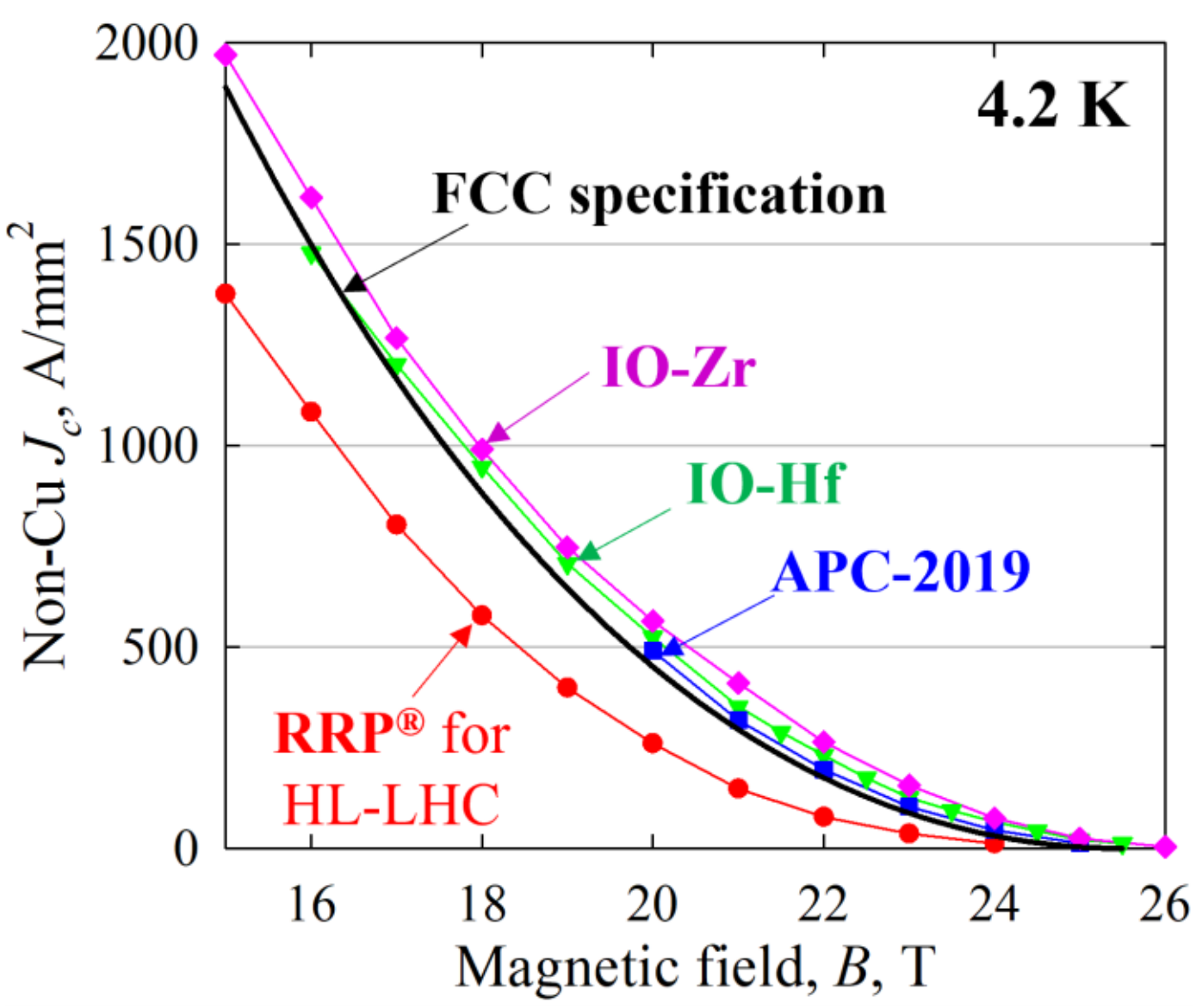

Figure 2. Non-Cu $J_c$s (4.2 K) of IO-Hf-705°C/65h, IO-Zr-700°C/60h, and APC-2019-700°C/70h, along with the FCC $J_c$ specification. Also shown are the results for an RRP® wire for the HL-LHC project [18].

It can be seen from Figure 2 that the non-Cu $J_c$s of both the IO-Hf and IO-Zr wires reach or surpass the FCC specification. The non-Cu $J_c$s of the IO-Zr wire are slightly higher than those of the IO-Hf wire and both are higher than those of the 2019 APC wire. The non-Cu $J_c$s of the 2019 APC wire shown here are lower than those reported in [10] for 0.84 mm diameter and 685°C/234h reaction, perhaps because the wire quality was better at 0.84 mm diameter and 685°C was better than 700°C for $J_c$. In general, we note that the APC wires have a higher advantage in $J_c$ at higher fields. For example, the non-Cu $J_c$s of the IO-Zr wire are 8% and 16% higher than the FCC specification (and 49% and 86% higher than the reference RRP® wire) at 16 T and 19 T, respectively. This is because APC wires have higher $B_{irr}$ than standard conductors [20,27], and the shift of the $F_p$-$B$ curve peak to a higher field [13-21] leads to enhanced $J_c$ at high fields (e.g., above 10 T) but reduced $J_c$ at low fields (e.g., below 5 T) [10].

It can also be seen that the recently-made IO-Hf and IO-Zr wires have much better electromagnetic stability than the APC wire made in 2019, despite their higher non-Cu $J_c$. For the 2019 APC wire the $I_c$ values below 20 T could not be measured due to quenches during *V-I* tests. The stability improvement is due to improved wire quality and RRR: the measured RRR values of the IO-Hf, IO-Zr, and the 2019 APC wire were 84, 91, and 22, respectively. The better RRR of recent APC wires are mainly due to the more conservative recipes. The average area fractions of residual Nb and fine-grain $Nb_3Sn$ layers within the filaments of the above wires were calculated from their SEM images (Figure 1) and are shown in Table 1. The unreacted Nb fractions for the IO-Hf and IO-Zr wires are higher than that of the 2019 APC wire, and are also much higher than those of the standard PIT wires, which are typically around 25% [28,29].

Table 1. Fractions of unreacted Nb and fine-grain $Nb_3Sn$ layers in the filaments for IO-Hf-705°C/65h, IO-Zr-700°C/60h, the 2019 APC wire, and the RRP® wire for HL-LHC.

| | IO-Hf-705°C/65h | IO-Zr-700°C/60h | 2019 APC wire: 700°C/70h | RRP® for HL-LHC |
|---|---|---|---|---|
| Unreacted Nb fraction, % | 35 | 34 | 28 | 10 |
| Fine grain fraction, % | 36 | 38 | 41 | 59 |

The fine-grain $Nb_3Sn$ layer $J_c$ values of these wires, which equal to the non-Cu $J_c$ values (Figure 2) divided by the fine-grain $Nb_3Sn$ area fractions (Table 1), were calculated and are shown in Figure 3. It is seen that the layer $J_c$ values of the IO-Zr and IO-Hf wires are similar, and are noticeably higher than the 2019 APC wire.

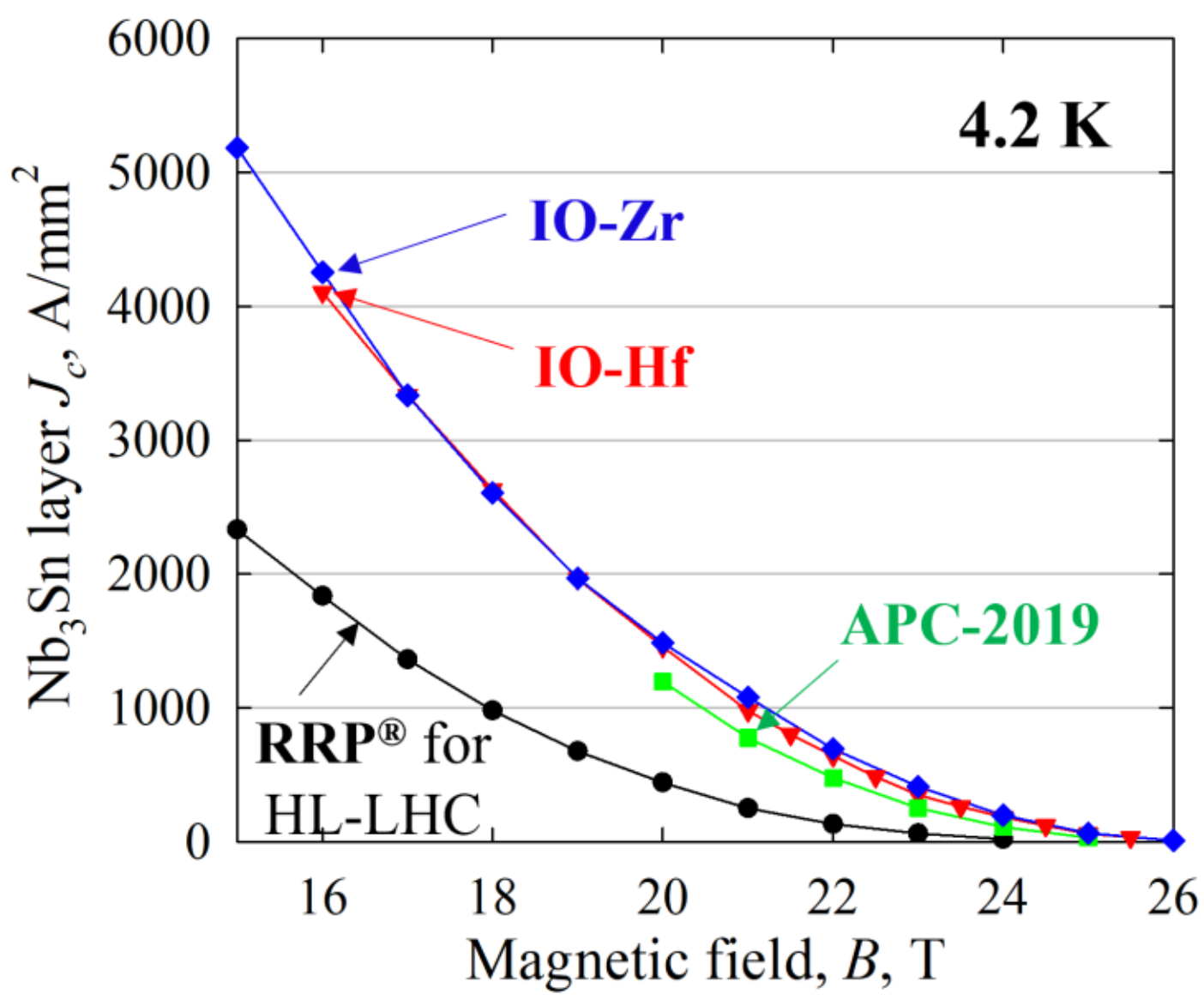

Figure 3. The $Nb_3Sn$ layer $J_c$ values for IO-Hf-705°C/65h, IO-Zr-700°C/60h, the 2019 APC wire, and the RRP® wire for HL-LHC.

In order to see if internal oxidation of Nb-Ta-Hf leads to different $B_{irr}$ and $B_{c2}$ values as compared to internal oxidation of Nb-Ta-Zr, the *R*-*B* curves of the above samples were measured and are shown in Figure 4. The $B_{c2}$ values at 10%, 50%, and 90% of the transitions are listed in Table 2. Also shown are the results for the RRP® control sample.

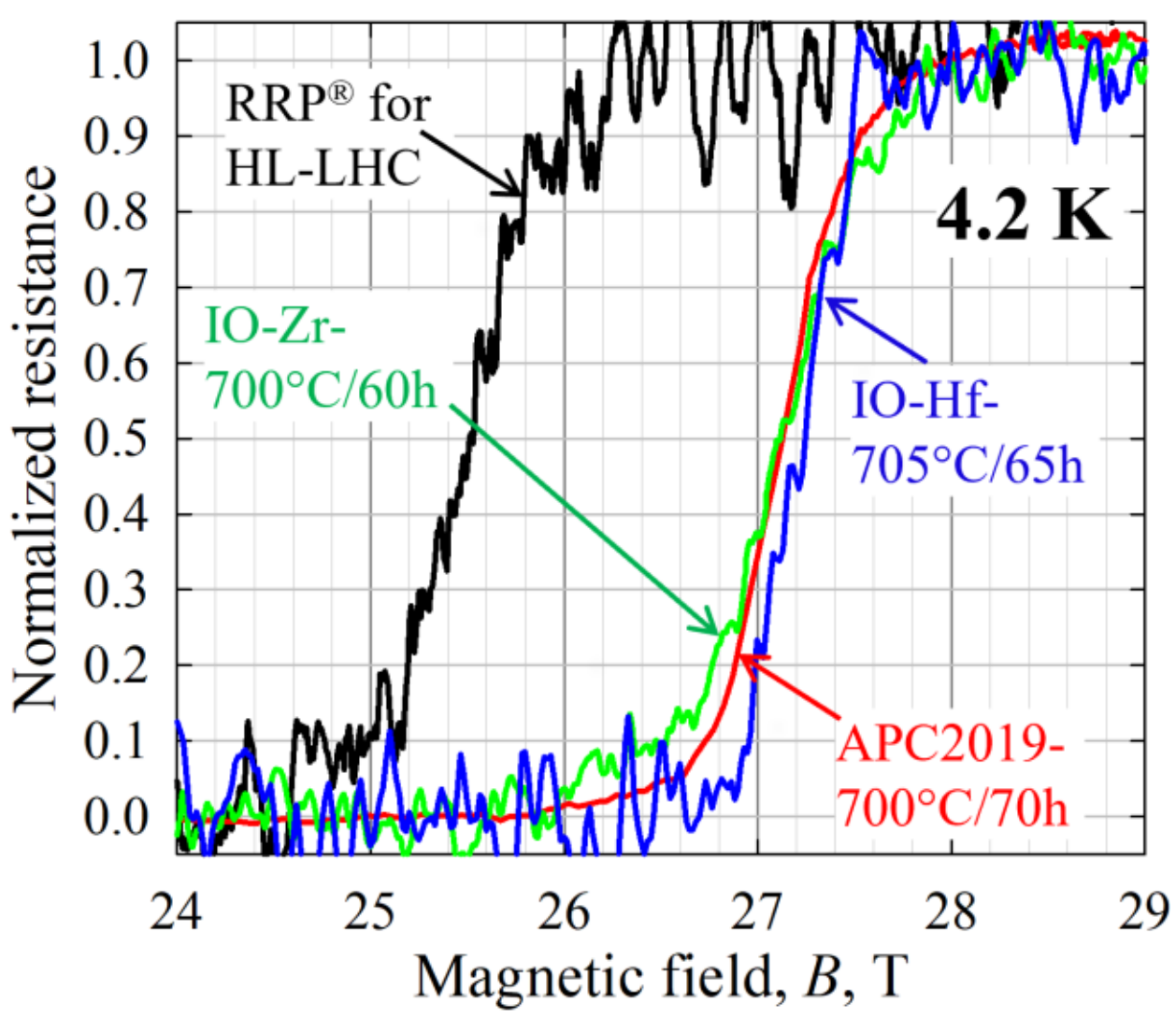

Figure 4. The *R-B* curves of IO-Hf-705°C/65h, IO-Zr-700°C/60h, the 2019 APC wire, and the RRP® wire for the HL-LHC.

Table 2. $B_{c2}$ (4.2 K) values as determined from the *R-B* curves

| | IO-Hf-705°C/65h | IO-Zr-700°C/60h | 2019 APC wire: 700°C/70h | RRP® for HL-LHC |
|---|---|---|---|---|
| $B_{c2}$-10%, T | 26.9 | 26.6 | 26.7 | 25.1 |
| $B_{c2}$-50%, T | 27.2 | 27.1 | 27.1 | 25.5 |
| $B_{c2}$-90%, T | 27.5 | 27.7 | 27.5 | 25.9 |

It can be seen that the $B_{irr}$ and $B_{c2}$ values for the APC wires based on internal oxidation of Nb-4at.%Ta-1at.%Hf and Nb-4at.%Ta-1at.%Zr were similar, and were 1.5-2 T higher than those of the RRP® wire. In fact, from Figure 2 it is seen that the $I_c$ values were still large enough to be measured (which means $I_c > 1$ A for the transport rig we used) for both the IO-Zr and the IO-Hf wires at 26 T but not for the RRP® wire at 24.5 T. In previous work we also measured $B_{irr}$ and $B_{c2}$ values for standard non-APC PIT wires [27], which were consistent with those measured by Godeke [29] and were about 1 T lower than those of our APC wires. A very high $B_{c2}$(4.2 K) of 29.2 T was reported in [20] for a $Nb_3Sn$ monofilament based on internal oxidation of Nb-4at.%Ta-2at.%Zr – of course the lower thermal strain in the samples studied in [20] might also contribute to their high $B_{c2}$ values. Our previous paper [27] gave a possible explanation for such a $B_{irr}$ and $B_{c2}$ increase in APC wires.

## 4. Discussion

It is seen from Figure 3 that the internal oxidation of Nb-4at.%Ta-1at.%Hf did not lead to higher layer $J_c$ than the internal oxidation of Nb-4at.%Ta-1at.%Zr, although our previous work demonstrated that the former led to somewhat more dramatic $F_p$-$B$ curve peak shift and grain refinement than the latter [24]. We still do not fully understand the reason for this yet. A possibility is that the diameters of the $HfO_2$ particles are mostly below 5 nm [24], smaller than the flux line core diameter, which equals to $2\xi$, where $\xi$ is the coherence length and is estimated to be ~3.5 nm for $Nb_3Sn$ at 4.2 K from the relation $B_{c2} = \Phi_0/(2\pi\xi^2)$ – in which $B_{c2}$ is 27-28 T (based on Figure 4) and $\Phi_0$ is the magnetic flux quantum ($2.07\times10^{-15}$ T·m$^2$). In contrast, the diameters of most $ZrO_2$ particles are in the 5-10 nm range, closer to $2\xi$, so each $ZrO_2$ particle may have higher flux pinning efficiency than each $HfO_2$ particle. On the other hand, the smaller $HfO_2$ particle size means higher number density per volume. Thus, a comprehensive model, which takes all microstructural factors (e.g., particle size, particle density, and $Nb_3Sn$ grain size) into consideration, is still needed to compare the flux pinning forces for the two scenarios.

From Figure 3 it is also seen that the IO-Zr wire has a higher layer $J_c$ than the 2019 APC wire, in spite of the fact that both were based on internal oxidation of an Nb-4at.%Ta-1at.%Zr alloy with similar Cu/Sn and O/Nb ratios. This is most likely because the IO-Zr wire has better wire quality and uniformity than the 2019 APC wire due to the use of a conservative recipe. After obtaining a large number of cross-sectional images we found that both the IO-Zr and the 2019 APC wire had some filaments with noticeably thinner $Nb_3Sn$ layers than the other filaments in some cross sections, but this problem was much severer in the 2019 wire. As examples, SEM images for three cross sections of the IO-Zr and three for the 2019 APC wire are shown in Figure 5, from which it is seen that the 2019 APC wire has more bad filaments than the IO-Zr, and most of the bad filaments are in the innermost layer of the filament array. The cause of this needs further investigation. We also found that such locally thin $Nb_3Sn$ layers were not due to compositional (e.g., Nb/Sn ratio) non-uniformity along the filament length, but typically due to Sn leakage into the surrounding Cu matrix during $Nb_3Sn$ layer growth, and that such a Sn leakage was usually caused by a local filament defect, such as an eccentric filament core or a filament distortion. A higher occurrence of such defects (and the associated Sn leakage) not only reduces RRR and stability, but also affects the transport $J_c$.

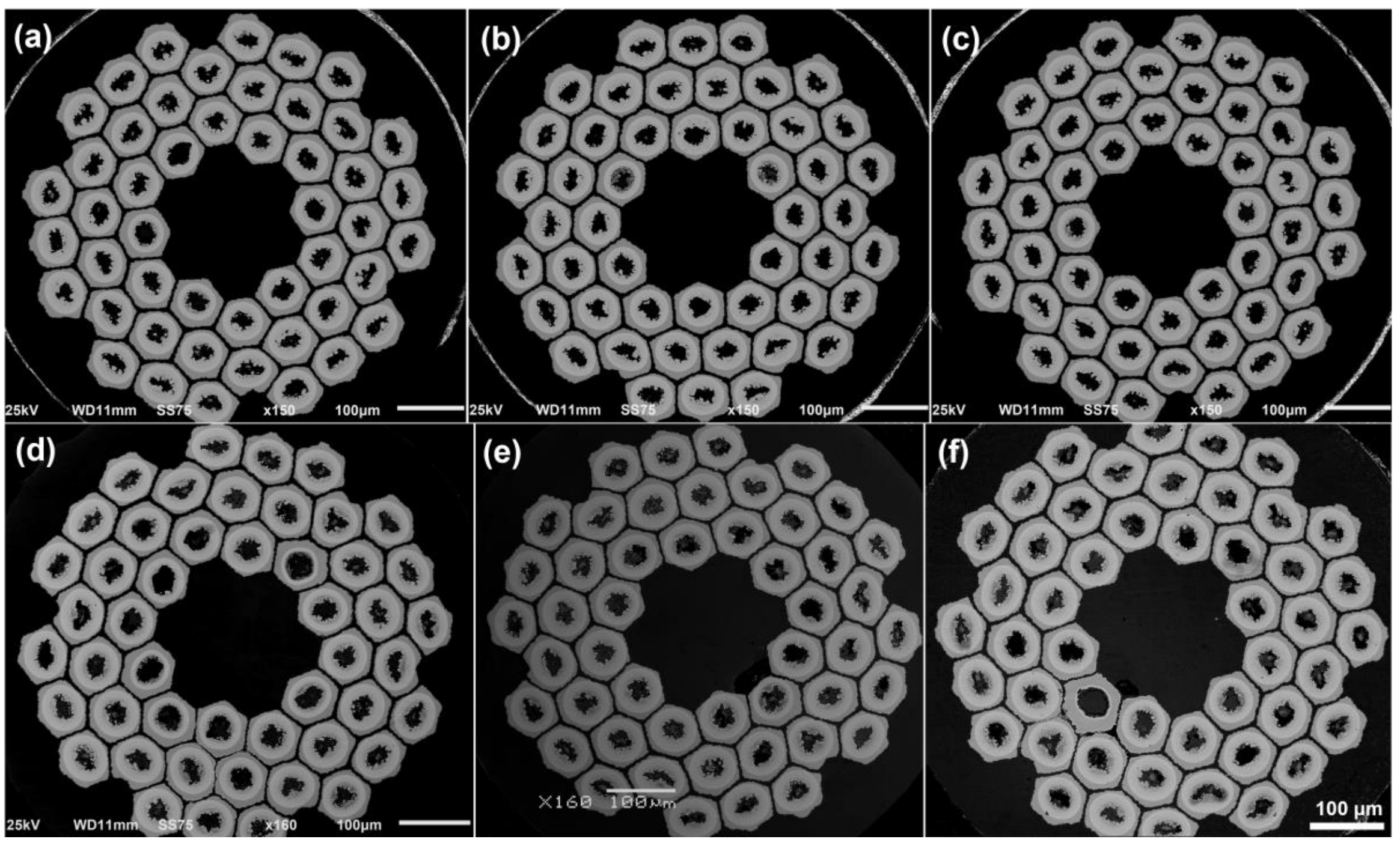

Figure 5. SEM images of (a)-(c) IO-Zr-700°C/60h and (d)-(f) the 2019 APC wire at various cross sections.

Although significant progress has been made in improving the electromagnetic stability of APC wires, further improvement is still needed before they can be practically used for making magnets. This can be achieved by several means. First, although the IO-Hf and IO-Zr wires have much better wire quality than the APC wires made in 2019, the fact that filament defects still exist in them (e.g., Figure 5) indicates that there is still room for further improvement. Improvement of wire quality requires further optimization of the wire recipe and design, as well as improvement of raw material quality and the wire fabrication process. Second, as mentioned earlier, the heat treatments used for these samples were not optimized. A further optimization in heat treatment is needed for further improvement of RRR and stability. Third, reduction of filament size, which has a significant influence on conductor electromagnetic stability [30], is also required. The IO-Hf and IO-Zr wires used in this work have filament sizes around 70 μm, which is still relatively large. Reduction of filament size requires increase of filament count in the wires. In 2019 we fabricated a 180/217-stack APC wire and drew it to 0.98 mm diameter (with a filament size around 50 μm) without any wire breakage, and its non-Cu $J_c$ reached the FCC specification at high fields [27]. However, because an aggressive recipe was used for that wire, it suffered the above-mentioned instability problem. Moving forward, we plan to make new 180/217-stack APC wires using more conservative recipes. Our near-term goal is to reduce the filament size to ~40 μm while increasing RRR to 150 and still keeping the non-Cu $J_c$ above the FCC specification. If so, we can expect that the electromagnetic stability will be significantly better than the IO-Hf and IO-Zr wires reported in this paper.

## 5. Conclusions

In this work we fabricated multifilamentary APC wires based on internal oxidation of Nb-7.5wt.%Ta-1at.%Hf and Nb-7.5wt.%Ta-1at.%Zr alloys. More conservative recipes were used relative to the APC wires made in 2019, leading to much better wire quality and higher RRR, and thus better electromagnetic stability. The conservative recipes did not lead to decrease in non-Cu $J_c$ in spite of a decrease in fine-grain $Nb_3Sn$ fraction, because of better filament quality and higher $Nb_3Sn$ layer $J_c$. The non-Cu $J_c$ values of both the IO-Hf and IO-Zr wires reach or surpass the FCC $J_c$ specification. The $Nb_3Sn$ layer $J_c$s of the IO-Hf were not higher than those of the IO-Zr, perhaps because the $HfO_2$ particle size is below the optimal flux pinning center size. Finally, there is still significant room for further improving the electromagnetic stability of APC wires by improving wire quality, optimizing heat treatment, and reducing filament size.

## Acknowledgements

This work was supported by the US Department of Energy through an Early Career Research Program Award and Hyper Tech SBIR DE-SC0017755. A portion of this work was performed in the NHMFL, which is supported by National Science Foundation Cooperative Agreement No. DMR-1644779 and the State of Florida. The authors want to thank the Applied Superconductivity Center (ASC) at the NHMFL for their help and for the use of their transport rig for the *V-I* tests. This manuscript has been produced by Fermi Research Alliance, LLC under Contract No. DE-AC02-07CH11359 with the U.S. Department of Energy, Office of Science, Office of High Energy Physics.

## References

[1]. Parrell J A, Field M B, Zhang Y Z and Hong S 2004 $Nb_3Sn$ Conductor Development for Fusion and Particle Accelerator *AIP Conf. Proc.* **711** 369-75

[2]. Field M B, Zhang Y, Miao H, Gerace M, and Parrell J A 2014 Optimizing $Nb_3Sn$ conductors for high field applications *IEEE Trans. Appl. Supercond.* **24** 6001105

[3]. Xu X 2017 A review and prospects for $Nb_3Sn$ superconductor development *Supercond. Sci. Technol.* **30** 093001

[4]. Abada A *et al.* 2019 FCC-hh: The Hadron Collider *Eur. Phys. J. Special Topics* **228** 755–1107

[5]. Benedikt M and Zimmermann F 2016 Status of the future circular collider study *CERN-ACC-2016-0331*

[6]. Ballarino A and Bottura L 2015 Targets for R&D on $Nb_3Sn$ Conductor for High Energy Physics *IEEE Trans. Appl. Supercond.* **25** 6000906

[7]. Tommasini D and Toral F 2016 Overview of magnet design options *EuroCirCol-P1-WP5 report* 4-6

[8]. Cooley L D, Ghosh A K, Dietderich D R and Pong I 2017 Conductor Specification and Validation for High-Luminosity LHC Quadrupole Magnets *IEEE Trans. Appl. Supercond.* **27** 6000505

[9]. Schoerling D 2019 The European 16 T Dipole Development Program for FCC and HE-LHC *U.S. Magnet Development Program workshop*, Batavia

[10]. Xu X, Peng X, Lee J, Rochester J and Sumption M D 2020 High Critical Current Density in Internally-oxidized $Nb_3Sn$ Superconductors and its Origin *Scr. Mater.* **186** 317-320

[11]. Benz M G 1968 The superconducting performance of diffusion processed $Nb_3Sn$ doped with $ZrO_2$ particles *Trans. of Met. Soc. of AIME* **242** 1067-70

[12]. Zeitlin B A, Gregory E, Marte J, Benz M, Pyon T, Scanlan R and Dietderich D 2005 Results on Mono Element Internal Tin $Nb_3Sn$ Conductors (MEIT) with Nb7.5Ta and Nb(1Zr+$O_x$) Filaments *IEEE Trans. Appl. Supercond.* **15** 3393-6

[13]. Xu X, Sumption M D, Peng X and Collings E W 2014 Refinement of $Nb_3Sn$ grain size by the generation of $ZrO_2$ precipitates in $Nb_3Sn$ wires *Appl. Phys. Lett.* **104** 082602

[14]. Hughes D 1974 Flux pinning mechanisms in type II superconductors *Phil. Mag.* **30** 293-305

[15]. Xu X, Sumption M D and Peng X 2015 Internally Oxidized $Nb_3Sn$ Superconductor with Very Fine Grain Size and High Critical Current Density *Adv. Mater.* **27** 1346-50

[16]. Xu X, Peng X, Sumption M D and Collings E W 2017 Recent Progress in Application of Internal Oxidation Technique in $Nb_3Sn$ Strands *IEEE Trans. Appl. Supercond.* **27** 6000105

[17]. Motowidlo L R, Lee P J, Tarantini C, Balachandran S, Ghosh A K and Larbalestier D C 2018 An intermetallic powder-in-tube approach to increased flux-pinning in $Nb_3Sn$ by internal oxidation of Zr *Supercond. Sci. Technol.* **31** 014002

[18]. Xu X, Rochester J, Peng X, Sumption M D and Tomsic M 2019 Ternary $Nb_3Sn$ conductors with artificial pinning centers and high upper critical fields *Supercond. Sci. Technol.* **32** 02LT01

[19]. Ortino M, Pfeiffer S, Baumgartner T, Sumption M, Bernardi J and Xu X 2021 Evolution of the superconducting properties from binary to ternary APC-$Nb_3Sn$ wires *Supercond. Sci. Technol.* **34** 035028

[20]. Buta F *et al* 2021 Very high upper critical fields and enhanced critical current densities in $Nb_3Sn$ superconductors based on Nb–Ta–Zr alloys and internal oxidation *J. Phys. Mater.* **4** 025003
[21]. Buehler C et al. 2020 Challenges and perspectives of the phase formation of internally oxidized PIT-type $Nb_3Sn$ conductors *IEEE Trans. Appl. Supercond.* **30** 6000805
[22]. Balachandran S, Tarantini C, Lee P J, Kametani F, Su Y, Walker B, Starch W L and Larbalestier D C 2019 Beneficial influence of Hf and Zr additions to Nb4at.%Ta on the vortex pinning of $Nb_3Sn$ with and without an O Source *Supercond. Sci. Technol.* **32** 044006
[23]. Tarantini C, Kametani F, Balachandran S, Heald S M, Wheatley L, Grovenor C R, Moody M P, Su Y F, Lee P J and Larbalestier D C. 2021 Origin of the enhanced $Nb_3Sn$ performance by combined Hf and Ta doping *Sci. Rep.* **11** 17845
[24]. Xu X, Peng X, Rochester J, Sumption M D, Lee J, Ortiz G A C and Hwang J 2021 The strong influence of Ti, Zr, Hf solutes and their oxidation on microstructure and performance of $Nb_3Sn$ superconductors *J. Alloys Compd.* **857** 158270
[25]. Takeuchi T, Asano T, Iijima Y and Tachikawa K 1981 Effects of the IVa element addition on the composite-processed superconducting $Nb_3Sn$ Cryogenics **21** 585-590
[26]. Tachikawa K and Sekine H 1982 Method for producing superconducting $Nb_3Sn$ wires U.S. Patent 4323402
[27]. Xu X, Sumption M D, Lee J, Rochester J and Peng X 2020 Persistent compositions of non-stoichiometric compounds with low bulk diffusivity: a theory and application to $Nb_3Sn$ superconductors *J. Alloys Compd.* **845** 156182
[28]. Segal C, Tarantini C, Lee P J and Larbalestier D C 2017 Improvement of small to large grain A15 ratio in $Nb_3Sn$ PIT wires by inverted multistage heat treatments *IOP Conf. Ser. Mater. Sci. Eng.* **279** 012019
[29]. Godeke A 2005 Performance boundaries in $Nb_3Sn$ superconductors *PhD dissertation* the University of Twente
[30]. Wilson M N 1983 Superconducting Magnets, Clarendon Press, Oxford